\documentclass[aps,prd,groupedaddress,nofootinbib,letterpaper]{revtex4}

\usepackage{amsmath}
\usepackage{amsfonts}
\usepackage{amssymb}
\usepackage{amsthm}

\usepackage{braket,mathtools}
\usepackage{mathrsfs}
\usepackage{stmaryrd}

\usepackage{hyperref}
\usepackage{color}

\newcommand{\cala}{{\cal A}}
\newcommand{\call}{{\cal L}}

\newcommand{\calh}{{\cal H}}

\newcommand{\beq}{\begin{equation}}
\newcommand{\eeq}{\end{equation}}

\newcommand{\eg}{{\it e.g.}}

\newcommand{\ue}{U_\epsilon}
\newcommand{\vx}{{\vec x}}

\begin{document}
\begin{titlepage}

\title{How is quantum information localized in gravity?}

\author{William Donnelly}
\email{donnelly@physics.ucsb.edu}
\affiliation{Department of Physics, University of California, Santa Barbara, CA 93106}

\author{Steven B. Giddings}
\email{giddings@ucsb.edu}
\affiliation{Department of Physics and Kavli Institute of Theoretical Physics, University of California, Santa Barbara, CA 93106}

\begin{abstract}

A notion of localization of information within quantum subsystems
plays a key role in describing the physics of quantum systems, and in particular is a {\it prerequisite} for discussing important concepts such as entanglement and information transfer.  While subsystems can be readily defined for finite quantum systems and in local quantum field theory, a corresponding definition for gravitational systems is significantly complicated by the apparent nonlocality arising due to gauge invariance, enforced  by the constraints.  A related question is whether ``soft hair" encodes otherwise localized information, and the question of
such localization also remains an important puzzle for proposals that gravity emerges from another structure such as  a boundary field theory as in AdS/CFT.  This paper describes different approaches to defining local subsystem structure, and shows that at least classically, perturbative gravity has localized subsystems based on a {\it split structure}, generalizing the split property of quantum field theory.  This, and related arguments for QED, give simple explanations that  in these theories there is localized information that is independent of fields outside a region, in particular so that there is no role for ``soft hair" in encoding such information.  Additional subtleties appear in  quantum gravity.  We argue that  localized information exists in perturbative quantum gravity in the presence of global symmetries, but that nonperturbative dynamics is likely tied to a modification of such structure.

\end{abstract}

\maketitle

\end{titlepage}

\section{Introduction}

A key question in quantum gravity is what mathematical structure underlies the theory.  This is important for understanding how quantum gravity matches onto the familiar structure of local quantum field theory (LQFT) in the weak-gravity limit.  It is also important for understanding the role of basic information-theoretic questions about the theory, which have been the subject of much recent discussion.  To take an example, there is much discussion about entanglement entropy.  However, as is emphasized for example in \cite{ZLS}, any entanglement structure is {\it relative to a subsystem structure}, and without specifying such subsystems, ambiguous.  Likewise, important questions surround {\it transfer of information}, particularly in trying to understand how black hole evolution can be unitary.\footnote{See for example \cite{SGmodels,BHQIUE,AMPS,NVU}.}  But, a clear description of transfer of information is also based on defining subsystems between which information is transferred; one needs to understand where information is located to describe how it is transferred.

While any subsystem structure is foundationally important, it is often taken for granted that such a structure exists, or, alternately, sometimes assumed that such a structure is emergent, {\it e.g.} from entanglement as in \cite{vanR,MaSu}.  In contrast, we feel it important to ask the question of whether giving a fundamental description of this structure -- which is poorly understood in gravity -- is an important aspect of arriving at a complete quantum theory of gravity.   In general, this question is closely tied to that of describing the nature of the degrees of freedom and localization properties of information of a theory, and the symmetries of a theory place  important constraints on such structure.

Indeed, in the LQFT limit, subsystem structure plays a foundational role in the theory.  In LQFT, subsystems are most precisely described\cite{Haag} in terms of subalgebras of the algebras of observables, associated with open regions of the spacetime manifold.  The key property of locality -- a cornerstone principle of the theory -- is specifically input as commutativity of these subalgebras.  

Such subsystem structure plays a key role in describing physics.  One example is in treating observation, where one would like a clean division between ``observer" and ``observee" subsystems.  In fact, it is hard to see how to give any complete description of physics without some such structure, providing such clean separations.  This point has been stressed by no less than Einstein, who for example in \cite{Eins}\footnote{See \cite{Howa}.} described the importance of this local {\it separability} in formulating and testing physical law.

Despite appearing to be a key concept, the question of subsystem structure is apparently remarkably subtle in gravity.  In finite quantum systems, subsystems are easily defined in terms of factorization of the Hilbert space into  tensor products.  As we will review, already in LQFT  this is problematic due to the infinite entanglement inherent in the continuum theory.  However, there one can base a definition on subalgebras, as just indicated.  But, quantum gravity -- if it matches onto perturbative general relativity in the weak field limit -- lacks an obvious subalgebra structure that corresponds with that of LQFT in the weak gravity limit, due to the need to gravitationally dress operators to satisfy the constraints\cite{DoGi1,DoGi2}.  

After reviewing these points, we will investigate another closely related notion of subsystem structure in LQFT, the notion of a {\it splitting} -- a kind of regularized factorization of the Hilbert space -- and then turn to the question of finding similar structures in gravity.  As a prelude, we first describe what we call ``split structures" in quantum electrodynamics (QED).  Here, the nonlocal structure of the gauge field does allows us to asymptotically determine information about the matter within a region, in a fashion similar to gravity.  However, as we show, the information that can be determined asymptotically is limited to that of the total electric charge.  This also addresses  the question of soft hair and charges in QED, which have been explored in this context of having asymptotic dependence on such local information\cite{HPS1,HPS2,Strorev}, and apparently provides a counterargument.  This raises the question of whether one can determine the information content of a region via measurements of its gravitational field outside this region.  We show that at least in the perturbative and classical theory, the answer is no,  in general one can only measure the Poincar\'e charges of a localized distribution, and this establishes the classical existence of such split structure.  Put differently, at least if one restricts to classical gravitational fields, there can be localized information in gravity, or localized ``gravitational qubits."  While the quantum situation is more subtle, and full treatment is deferred for future work, we also see that if global symmetries are present, these can  result in such split structure -- or ``local qubits" -- in the perturbative theory.  We close by discussing the non-perturbative situation, where arguments have been given for an origin of ``gravitational holography\cite{Maroholo1,Maroholo2,Jaco}" in the constraints.
There, despite finding a close link between unitary evolution and gravitational nonlocality, we do not yet find an independent description of either.

\section{Approaches to defining subsystems}

Given the importance of defining subsystem structures in order to describe physical processes and their observation, we would like to understand how to think about subsystems in a quantum theory of gravity.  We begin by recalling  analogous definitions for simpler systems.

\subsection{Tensor factorization}

For quantum systems with finite-dimensional Hilbert spaces, a standard definition of subsystem structure is given by providing a tensor factorization of the Hilbert space, \eg, 
\beq\label{hfact}
\calh=\calh_1\otimes\calh_2\ ,
\eeq
with the tensor factors $\calh_1$ and $\calh_2$ corresponding to the distinct subsystems.  Then, for example, notions such as entanglement of joint states of these subsystems, or information transfer between them, can be simply defined with respect to this subsystem decomposition.  The same kind of structure can be used for locally-finite systems, \eg\ lattice models where we can think of associating factor Hilbert spaces to lattice sites.

In continuum quantum field theory, the problem of defining subsystems becomes more subtle. 
 In general, if we consider two adjacent spatial regions, the Hilbert space does not correspondingly factorize as in \eqref{hfact}.  
This is a generic property of continuum field theories;\footnote{See \cite{BuVe}, and  for  reviews and further references \cite{Haag,Yngv}; also see \cite{Requ}.} while technically it is known as the type-III property of the von Neumann algebras acting on their Hilbert spaces, practically it can be understood as arising from the entanglement of adjacent degrees of freedom arbitrarily deeply into the ultraviolet, in states that have ultraviolet behavior like that of the vacuum.  
Conversely, an attempt to define unentangled states of the form \eqref{hfact} for adjacent regions produces infinite-energy states resulting from `breaking' all this entanglement. 
(Such states have been seriously considered as physical states for black holes, and called ``firewalls", in \cite{AMPS}.)
This failure of factorization is directly seen in the generic UV divergences that are found in quantum field theory (QFT) when computing the entropy of any region.  

\subsection{Commuting subalgebras}

To avoid this problem, a QFT definition of subsystems can instead be based on the structure of the algebra of observables, as in \cite{Haag}.  Specifically, associated to each region of spacetime is a subalgebra of observables supported on that region, and the subalgebras associated to spacelike-separated regions commute.  These commuting subalgebras can be though of as providing a  definition of subsystems; notice that in the finite case the decomposition \eqref{hfact} also implies the existence of such subalgebras.

The problem of defining subsystems becomes considerably more subtle for gravity.  While we do not presently know the full description of the degrees of freedom of gravity, in the low-energy limit this should match onto a weak-field description given by perturbative general relativity, whose gauge symmetries are the diffeomorphisms.  This theory shares with more typical QFTs the problem of defining a Hilbert space factorization \eqref{hfact}.  While one doesn't expect the weak-field description of gravity to be valid arbitrarily far into the UV, there do not appear to be any obvious clues from the UV structure of gravity about how to define such a Hilbert space factorization.  And, when one instead considers the algebraic approach, 
one sees that the problem is worse than that of field theory.  Here, unlike in QFT, gauge invariant observables cannot be localized within definite regions of spacetime, and so one does not have an obvious definition of subsystems via commuting subalgebras.  

This behavior can be seen directly by working perturbatively in Newton's constant $G$, in the weak-field regime.  Operators that would be localized in the QFT description -- such as a scalar field operator $\phi(x)$ -- must be ``gravitationally dressed" in order to become gauge invariant\cite{SGalg,DoGi1,DoGi2}.  Once dressed, the support of the resulting operators generically extends to infinity, as shown in \cite{DoGi2}.   Such operators will not commute at spacelike separation, obstructing a definition of subsystems associated to finite neighborhoods of spacetime.  Indeed, there are arguments that the problem is even worse\cite{SGalg,DMM}, since there is even an apparent obstacle to restricting the gravitational dressing to definite infinite neighborhoods extending to infinity.  

This problem appears to be present in {\it any} proposed theory of gravity that matches onto general relativity in the weak field limit, and thus these statements appear to be highly constraining.  Very plausibly, these aspects of the Hilbert space structure and/or algebraic structure are intrinsic to the fundamental structure of quantum gravity, and thus need to be better understood.  Alternately, one might consider the possibility that quantum gravity emerges from a more basic structure that behaves like local quantum field theory, or something very similar to it.  But, in that case, a key question is how the kind of algebraic structure just described could emerge in the weak-field or long distance limit; this in fact appears to be a strong set of constraints on any putative theory of emergent gravity.  (See \cite{SGpots}, and for closely related comments \cite{Marolf:2014yga}, and references therein.)

\subsection{Split structures}
\label{splitst}

If an algebraic definition of subsystems of gravitational systems cannot be given that matches onto the QFT definition of a subsystem in the weak-field limit, an alternative to investigate is whether there is a weaker mathematical structure that provides a definition sufficient for purposes of physics.  For example, one problem with an algebraic definition stems from the fact that a high power of an operator creating one particle can create many particles, which will then have a strong gravitational field spanning an increasingly large region.  A possible weaker definition of a subsystem structure is to specify instead a set of states that differ within a region, but which are indistinguishable by observables outside that region.  This does not necessarily imply either a tensor or algebraic factorization, but does imply that information outside the region is independent of information contained within the region. 

We first outline the description of such structure in a non-gauge field theory with field $\phi$ and Hilbert space $\calh$.  This description is based on the split property\footnote{See, {\it e.g.}, \cite{Haag}, and references therein.} of the operator algebras associated with sufficiently well-behaved field theories.  Consider a spatial neighborhood $U$ in flat spacetime; there are commuting subalgebras $\cala_U$ and $\cala_{U'}$ associated with this neighborhood and its complement.  However, as noted above, this does not imply a factorization of the Hilbert space.  For example, the vacuum $|0\rangle$ has infinite entanglement of degrees of freedom arbitrarily close to the boundary of $U$.\footnote{In fact the type III property is much stronger than simply infinite vacuum entanglement.
More precisely, the algebra of observables in a finite region in quantum field theory is believed to be a type III${}_1$ von Neumann algebra.
For \emph{any} state on this algebra the modular operator, which in the algebraic setting plays the role of the density matrix, has a continuous spectrum that includes all positive real numbers.}

But, suppose we are given a ``thickening'' of $U$, $U_\epsilon$, which extends it outward by a characteristic distance $\epsilon$.  Then, the split property tells us that there is a state where degrees of freedom in $U$ and $U_\epsilon'$ are disentangled.  Specifically, there is a state $|U_\epsilon\rangle$ such that for $x\in U$ and $x\in U_\epsilon'$, 
\beq
\langle U_\epsilon| \phi(x)\phi(x') |U_\epsilon\rangle = \langle0| \phi(x)|0\rangle \langle0|\phi(x') |0\rangle\ ,
\eeq
with this relation extending to more general operators supported in $U$ and $U_\epsilon'$.
The state $|U_\epsilon\rangle$  may be written in terms of a unitary operator ${\cal U}(U_\epsilon)$ acting on the vacuum, 
\beq\label{fsplit}
|U_\epsilon\rangle = {\cal U}(U_\epsilon)|0\rangle\ ;
\eeq
for free fields ${\cal U}(U_\epsilon)$ is expected to be a  Bogoliubov transformation.   
Then, consider states obtained by acting with the operators in $\cala_U$ on $|U_\epsilon\rangle$.  
These will be indistinguishable by operators of $\cala_{U_\epsilon'}$.  
Specifically, if $|\psi,U_\epsilon\rangle$ and $|\tilde\psi,U_\epsilon\rangle$ are two such states, and $A$ is an operator in $\cala_{U_\epsilon'}$, then 
\beq
\langle\tilde\psi,U_\epsilon|A|\psi,U_\epsilon\rangle = \langle\tilde\psi,U_\epsilon|\psi,U_\epsilon\rangle \langle0|A|0\rangle\ .
\eeq
  Thus we have a Hilbert space $\calh_{U_\epsilon}\subset \calh$ of states associated to $U_\epsilon$ that ``look just like vacuum" to an observer outside $U_\epsilon$.\footnote{
The split property itself is somewhat stronger, and says there is a Type I factor with $\cala_U \subset \mathcal{F} \subset \cala_{U_\epsilon}$. 
In other words, there is a factorization $\calh = \calh_1 \otimes \calh_2$ with $\cala_U \subset \mathcal{B}(\calh) \otimes I$ and $\cala_{U_\epsilon'} \subset I \otimes \mathcal{B}(\calh_2)$.}

One of the goals of this paper is to investigate whether such a weaker notion of subsystem -- based on a this kind of Hilbert-space structure -- is possible in gravity.  More colloquially, this is a question of whether there can be localized qubits in a gravitational theory -- that is, can we associate some information to a given region (qubit in ``up" or ``down" state) without that information being accessible outside that region?  

 We will initially explore this question in the weak-field, perturbative, limit, by investigating what information is available to an observer outside a given region who has access to field observables, including those of the gravitational field, outside the region in question.  

The discussion of this paper has an apparent connection with the soft-hair story, being pursued by Strominger and others\cite{HPS1,HPS2,Strorev}.  Specifically, that story investigates an infinite number of conserved charges present at infinity, in either gauge theory or gravity.  An important question, then, is whether knowledge of all those charges implies complete knowledge of the state; if so, that is an important constraint, which may for example help with the unitarity problem for black holes.  These soft charges are examples of quantities that may be measured outside a region, in an attempt to determine the state inside.

We next make the preceding discussion more concrete, working perturbatively in a weak field expansion, or expansion in $G$.  We can illustrate some of the considerations by first comparing with the case of QED.  

\section{Subsystems in QED}

\subsection{Algebraic definition}

Suppose that we consider a spatial neighborhood $U$, and wish to define a subsystem associated to that neighborhood.  In QED, there is a subalgebra of the algebra of observables that creates excitations localized in $U$.  For example, consider the electron field operator $\psi(x)$.  This is not gauge-invariant, and must be dressed to provide a gauge-invariant operator.  There are many choices of dressing -- essentially differing by choices of the radiation (source-free) electromagnetic field; one choice is 
\beq\label{infwils}
\Psi_\Gamma(x)= e^{iq\int_\Gamma A}\psi(x)
\eeq
where $\Gamma$ is a curve running from $x$ to infinity.  This operator will be invariant under gauge transformations,
\beq
\psi(x)\rightarrow e^{-iq\Lambda(x)}\psi(x)\quad,\quad A\rightarrow A- d\Lambda
\eeq
for $\Lambda(x)$ with compact support.  $\Psi_\Gamma(x)$ creates an electron, together with an electric ``Faraday line" running along $\Gamma$.\footnote{While this field configuration is singular, due to restriction of the field lines to a single curve $\Gamma$, one may regularize by ``thickening" into a small tube surrounding $\Gamma$.}  Thus $\Psi_\Gamma(x)$ has noncompact support, and in general will not commute with other operators outside $U$.  However, a subalgebra of observables restricted to $U$ can be found by considering instead operators of the form
\beq \label{wilsxx}
W(x,x')={\bar \psi}(x') e^{iq\int_\Gamma A}\psi(x)\ ,
\eeq
where $x$, $x'$, and $\Gamma$ (or, its thickened version) are contained in $U$.  
We can consider the subalgebra of operators with support restricted to $U$, {\it e.g.} generated by sums
and products of such operators, and the  action of the operators of this subalgebra on the vacuum $|0\rangle$ will create excitations which we can think of as ``localized to $U$."  This provides an algebraic  definition of a localized subsystem in QED.  

\subsection{Split structure}

We may alternately describe subsystems in terms of split structure on the Hilbert space, which also extends to QED\cite{HoDa}.
The QED case provides a paradigm whose extension to gravity we wish to explore, and also makes connection with recent discussions\cite{HPS1,HPS2,Strorev} of ``soft hair" in QED.  

In line with the discussion of section \ref{splitst}, given a neighborhood $U$, we expect to be able to use a splitting associated to an extended neighborhood $U_\epsilon$ to describe a Hilbert space of states localized to $U$, which cannot be distinguished by observables outside $U_\epsilon$.  Specifically, if $\cala_U$ is the algebra of operators associated with $U$ in the $e=0$ theory, we can consider the algebra $\hat\cala_U$ of operators in the $e\neq0$ theory which are dressings of the operators in $\cala_U$.  These will not necessarily be localized to $U$, an example being \eqref{infwils}.  But we can also define the subalgebra $\hat\cala^R_U$ of operators -- such as $W(x,x')$ in \eqref{wilsxx} -- whose support is restricted to $U$.

Let us likewise define $\hat\cala^R_{U_\epsilon'}$ as the algebra of dressed operators with support restricted to $U_\epsilon'$.  Then, if $|U_\epsilon\rangle$ is the split vacuum associated to $U_\epsilon$, we have
\beq \label{qedsplit}
\langle U_\epsilon|AA'|U_\epsilon\rangle = \langle 0| A|0\rangle \langle 0| A'|0\rangle 
\eeq
for $A\in \hat\cala^R_U$ and $A'\in \hat\cala^R_{U_\epsilon'}$.  In particular, this means that the collection of states created by $A\in \hat\cala^R_U$ acting on $|U_\epsilon\rangle$ are indistinguishable by measurements made with dressed operators outside $U_\epsilon$.  

It is also important to consider elements in $\hat \cala_U$ whose dressing extends to infinity.  Acting on $|\ue\rangle$, these create states which {\it can} be distinguished outside $\ue$ -- for example, the presence of the dressed operator \eqref{infwils} can be diagnosed by its nontrivial commutator with the field strength operator $F_{\mu\nu}$ which detects the field lines running out to infinity.  This raises the interesting question of how much information can be detected about states in $U$ by measurements conducted outside $U_\epsilon$.  Can we, for example, detect the configuration of a collection of charged particles inside $U$?

The answer to the latter question is {\it no}; in general we can only detect the total charge.  To see this, consider an undressed operator that is a product $\prod_i \psi(x_i)$.  This operator can be dressed as in \eqref{infwils} with a collection of Faraday lines running from the points $x_i$ to a common point $x_0\in U$. The resulting operator is not gauge invariant at $x_0$, but that can be repaired by dressing the point $x_0$ with an operator extending to infinity; this could be a Faraday line like in  \eqref{infwils} but with total charge matching that of the $\psi(x_i)$'s, or alternately could be a dressing that creates a Coulomb field with the corresponding total charge -- for an explicit formula, see eq.~(15) of \cite{DoGi1}.  Clearly the argument extends to a product of $\psi$'s and $\bar \psi$'s.  So, an arbitrary operator in $\cala_U$ -- creating a configuration of charged particles -- corresponds to a dressed operator in $\hat\cala_U$ whose configuration outside $U$ only depends on the total charge; we can also alter the part of the dressing inside $U$ to produce even more operators indistinguishable outside $U$. Put differently, for arbitrary operators in $\cala_U$, there is an algebra $\hat \cala_U^S$ of ``standardly dressed" operators, whose dressing outside $U$ only depends on the charge $Q$ of the operator.  If $A_\alpha$, $A_\beta$ are two such operators with charge $Q$, and $A'\in \hat\cala^R_{U_\epsilon'}$, then 
\eqref{qedsplit} becomes 
\beq\label{Qsplit}
\langle U_\epsilon|A_\beta^\dagger A'A_\alpha |U_\epsilon\rangle = \langle 0| A_\beta^\dagger A_\alpha|0\rangle \langle Q| A'|Q\rangle 
\eeq
where the state $|Q\rangle$ depends only on the standard choice of dressing outside $U$ ({\it e.g.} Faraday, Coulomb, or other).  Thus, for any given $Q$ there is an infinite Hilbert space of states localized to $U$, carrying that charge, which are indistinguishable outside $U$.\footnote{One may alternately describe such states by acting with operators of $\hat \cala^R_U$ on a state of a given charged sector, deriving a statement like \eqref{qedsplit}; one can check that a dressing such as a Faraday line does not impede this.  Also, notice that such states of course do not remain indistinguishable outside $U$; {\it e.g.} an initial dipole connected by a Faraday line will radiate as it transitions to a dipole field configuration.}  In this sense, one also has subsystems that carry a charge.  
\subsection{Relation to ``soft hair"}

There has been considerable recent discussion of the possibility that asymptotic gauge or gravitational fields carry information about a localized state of particles, particularly through the existence of conserved ``soft charges;" for a review with references see \cite{Strorev}.  The preceding discussion sheds some general light on the association between such asymptotic fields and the information content of a region, in QED.

Specifically, we have just seen that for a neighborhood $U$ and any given charge $Q$, there is an infinite Hilbert space of states which are indistinguishable by measurements of the electromagnetic dressing outside $U$.  This means there is an infinite amount of information that can be ``stored in $U$" which is inaccessible outside.  The soft charges are of the form
\beq \label{softc}
\lim_{r\rightarrow\infty} r^2 {\hat r}^i E_i(r \hat r^j)\ ,
\eeq
{\it i.e.} the asymptotic configuration of the radial electric flux.  Clearly this is a special case of more general expressions involving the field operator $F_{\mu\nu}$ outside $U$, which are insensitive to the states of this Hilbert space.

There is a lot of information contained in the charges \eqref{softc}, and more generally in the configuration of the field $F_{\mu\nu}(x)$ outside $\ue$.  To understand what this information describes, notice that the soft charges change if we change the above standard choice of dressing, {\it e.g.} from a Faraday line to the Coulomb dressing.  Such a change of dressing corresponds to a change in the {\it radiative} part of the electric field -- their difference is a field that is a homogeneous solution of Maxwell's equations\cite{PFS,DoGi1}.  So, for a given collection of states with one dressing, we can add a radiative field to produce another dressing.  Evolved backward in time, this maps to different initial data on ${\cal I}^-$ corresponding to choices of different incoming radiative electromagnetic fields -- but is {\it not} necessarily associated with the precise configuration of the matter sourcing this field, aside from the information in the total charge.

So, in short, this demonstrates that the ``soft hair" of QED {\it is not} associated with the state of matter in a region, aside from its total charge, but {\it is} associated with the choice of radiative electromagnetic field that has been superposed on that matter configuration.\footnote{Related comments have been made previously in algebraic quantum field theory in \cite{Bucholz1982}, and in the context of the S-matrix in \cite{MiPo}.}  We can store an arbitrary amount of information in ``electrically-charged qubits" within a region (including inside a black hole), without that information being simultaneously accessible in the electromagnetic field outside.

\section{Subsystems in Gravity?}

\subsection{Algebraic approach}

We now turn to analogous considerations for gravity, for simplicity coupled to a single scalar field $\phi(x)$.  The operator $\phi(x)$ is not gauge-invariant, and must be dressed to provide a gauge-invariant operator.  Again, there are many choices of dressing -- essentially corresponding to different choices of the radiation (source-free) gravitational field.  Working to leading order in $\kappa=\sqrt{32\pi G}$, one choice was given in \cite{DoGi1},
\beq\label{phifar}
\Phi_\Gamma(x)= \phi(x^\mu+V^\mu_\Gamma)\ ,
\eeq
where $V^\mu_\Gamma$ is an integral of an expression involving the metric over a curve $\Gamma$ (taken in \cite{DoGi1} to be parallel to the $z$ axis) running from $x$ to infinity.\footnote{As in the QED case, such a line may be regularized by ``thickening" it.  In gravity it can be particularly helpful to thicken to a small but finite angle.  In fact, a particularly natural choice is a dressing thickened to subtend a small finite angle within a two-plane that extends to infinity.}  This operator is invariant (to leading order in  $\kappa$) under gauge transformations
\beq
\delta\phi = -\kappa\xi^\mu\partial_\mu\phi \quad ,\quad \delta g_{\mu\nu} = -\kappa \call_\xi g_{\mu\nu}
\eeq
with $\xi^\mu(x)$ of compact support.
This operator creates a scalar particle, together with a gravitational flux running along $\Gamma$.
Thus $\Phi_\Gamma$ has noncompact support, and in general will not commute with other operators outside a region $U$ containing $x$.

But now in gravity there is no direct analog of the localized operators $W(x,x')$ in \eqref{wilsxx}.  In fact, the {\it Dressing Theorem} of  \cite{DoGi2}, prohibits the existence of such localized operators.  This theorem shows that for {\it any} quantum field operator localized to $U$, the dressing at leading order in $\kappa$ must extend to infinity.  The theorem formalizes the more colloquial statement that one cannot screen the gravitational field.  And, it precludes an algebraic definition of localized subsystems:  one does not find a subalgebra of operators localized to $U$ which commutes with all operators outside $U$.  In fact, it appears that the situation is worse\cite{SGalg,DMM}:  while the operator $\Phi_\Gamma$ is localized to a narrow neighborhood surrounding $\Gamma$ which extends to infinity, we expect that in full nonlinear gravity, $\Phi_\Gamma^N$, for large $N$, affects a region that expands with $N$.\footnote{This behavior would be interesting to explore further.}

Thus, gauge-invariant operators in gravity are intrinsically delocalized, and there is no clear algebraic notion of subsystems which matches onto that for LQFT in the $G\rightarrow0$ limit.\footnote{Of course, if one had a description of the full Hilbert space for quantum gravity, one could {\it formally} define tensor factorizations or operator subalgebras yielding formal subsystems.  However, in general there is no reason for such a definition to match the familiar definitions of subsystems that we use in LQFT in the $G\rightarrow0$ limit, and in particular to respect the localization properties of LQFT, suggesting limited utility for such formal definitions.}  
This suggests looking for a weaker notion of subsystems in gravity.  It is natural to inquire whether such a notion can be based on an analog of the split structure described in the preceding sections.

\subsection{Split structure}

Specifically, let $\calh$ be the full Hilbert space of the theory.  
If we choose a neighborhood $U$, we can ask if there are Hilbert subspaces $\subset \calh$ of states which are indistinguishable outside $U_\epsilon$.
Note that in in order to discuss such a construction, we work perturbatively about a fixed background metric.

From the preceding discussion, we already see that gravity has significant differences from QFT.  Specifically, states $|\psi,U_\epsilon\rangle$ like those of section \ref{splitst}, once dressed, will not produce vacuum correlators for observables outside $U_\epsilon$.  There are two reasons for this.  The first is the Dressing Theorem: operators that would be localized to $U$ in QFT must have dressing extending to infinity, and this dressing will have measurable effect on asymptotic observables.  Moreover, even the ``fundamental'' split state $|U_\epsilon\rangle$ in the $G=0$ theory (see \eqref{fsplit}) is expected to produce measurable effect on asymptotic observables, when generalized to a state $|\widehat U_\epsilon\rangle$ at $G\neq0$ which satisfies the constraints.
This is because $|U_\epsilon\rangle$ differs from the vacuum, hence has energy-momentum, hence leads to a non-trivial asymptotic gravitational field once dressed.

This means that in gravity we inevitably face a problem analogous to that of the charged subsystems in QED -- excitations in a given neighborhood will generically produce nontrivial asymptotic fields, arising from the energy-momentum distribution of the state.  

We can then ask whether there is an analogous definition of gravitational subsystems, arising from a {\it split structure} similar to that in \eqref{Qsplit}.  Given the algebraic problems arising from high powers of operators, it seems better to base this directly on the structure of the Hilbert space, rather than attempting to define it algebraically.  Specifically, given a $U_\epsilon$, we can ask whether there are Hilbert spaces $\calh_{U_\epsilon}^i\subset \calh$ such that observables outside, while they can distinguish states $|\psi,U_\epsilon\rangle\in \calh^i_{U_\epsilon}$ from  vacuum, cannot distinguish states in $\calh^i_{U_\epsilon}$ from each other.  Specifically, for any operator $A'$ localized outside $U_\epsilon$, and for any two states $|\psi,U_\epsilon\rangle,|\tilde\psi,U_\epsilon\rangle  \in \calh^i_{U_\epsilon}$,
\beq\label{grsplit}
\langle\tilde\psi,U_\epsilon|A'|\psi,U_\epsilon\rangle = \langle\tilde\psi,U_\epsilon|\psi,U_\epsilon\rangle \langle i|A'|i\rangle ; 
\eeq
compare eq.~\eqref{Qsplit}.   In the expression on the right, the matrix element of $A'$ can be non-trivial, but only depends on the label $i$ of the Hilbert space $\calh_{U_\epsilon}^i$, and not on which states have been chosen from $\calh_{U_\epsilon}^i$.  In the case of QED, the label $i$ was the charge $Q$.  In gravity,  at least classically we expect the label $i$ might include the total energy contained in the region; this total energy could be measured via the exterior metric, but we could ask whether there are multiple states that are indistinguishable by such total energy measurements.  Of course, other measurements involving the exterior metric can also be made, and so it is a nontrivial question whether there are {\it any} such nontrivial gravitational split structures.  If there are, they will not just be characterized by their energy.

To investigate the possible existence of such split subsystems in perturbative gravity, we take an approach like that for QED.  
Given the neighborhood $U$ and a non-gravitational ($G=0$) LQFT, there is a subalgebra $\cala_U$ of operators with support in $U$ that commute with the subalgebra associated with the exterior of $U$.  If this theory is then coupled to gravity, the operators in $\cala_U$ must be gravitationally dressed in order to be gauge-invariant.  As in QED, this dressing is not unique.  Moreover, not all operators in $\cala_U$ have a consistent perturbative dressing; for example, some operators will create states whose energy exceeds a bound given by the linear size $R$ of $U$, $GE\gtrsim R$, and clearly are associated with non-perturbative gravitational fields.\footnote{This bound is the {\it locality bound} of \cite{GiLia,GiLib,LQGST}, and in $D$ spacetime dimensions becomes $GE\gtrsim R^{D-3}$.} Let $A_U$ be such a set of perturbatively-dressed operators that do not create strong fields.  Then, we can act on a fundamental split state $|\widehat U_\epsilon\rangle$ with any combination of such operators that does not produce a strong  field.  This set of states gives a candidate Hilbert space of ``weak gravitational states'' associated to $U_\epsilon$.  We can  ask whether this Hilbert space decomposes into smaller Hilbert spaces $\calh_{U_\epsilon}^i$ that are indistinguishable outside $U_\epsilon$ in the sense of \eqref{grsplit}.   If so, this would show that information can be localized ``inside $U$,'' in a way that cannot be registered outside  $U_\epsilon$ -- defining a notion of a split subsystem in gravity.

We would na\"\i vely expect that at a minimum, the labels $i$ include the different possible gravitational charges carried by a subsystem; these are the momenta $P_\mu$ and angular momenta $M_{\mu\nu}$, generating the Poincar\'e algebra.  These quantities are measurable at infinity.  An important question is whether there is such a division of $\calh$ into split subsystems, and what the full set of labels $i$ is.  If there is no such division, that indicates that {\it all} states can be distinguished by measurements outside the region $\ue$, and thus that information is in this sense delocalized in gravity.  Checking this structure is more challenging for gravity than for QED; we first turn to a discussion of the dressing of classical matter.

\subsection{Classical subsystems in gravity}

As a prelude to the quantum problem, this subsection will address the question of the analogous classification of classical subsystems in gravity.  Specifically, we consider the following question.  Given a neighborhood $U$, are there classical particle configurations in $U$ that can be consistently gravitationally dressed, and for which that dressing does not allow us to distinguish different configurations?  We will show that the answer is yes, and that, analogous to the case of QED, the dressing for a given configuration may be put in a standard form that depends only on the Poincar\'e charges of the configuration.  

Specifically, consider a classical energy-momentum distribution $T_{\mu\nu}$, localized to the neighborhood $U$.  For simplicity, we represent this as a sum over a discrete collection of point sources.\footnote{As a technical point, we take the region $U$ to be star-shaped: we assume that there exists a point, taken to be the origin, such that the straight line connecting any point in $U$ to the origin lies within $U$.
This is because the gravitational Wilson line solutions we will construct are most naturally defined on geodesics. 
While we see no obstacle to defining gravitational Wilson lines supported on non-geodesic curves, we expect this to add additional technical complication that we wish to avoid.}
The question is whether for such a configuration, we can consistently find a (perturbative) gravitational field, whose behavior outside $U$ just depends on the Poincar\'e charges of the collection, similar to the QED case.  If so, we have classical subsystems corresponding to different configurations of particles with the same total charges.

The answer to this question is yes, as may be shown by a similar construction to that for QED.  An essential role is played by the constraints, which are the gravitational analog of Gauss' law, and must be satisfied by consistent data on a given time slice.
Taking our initial data surface to be the $t = 0$ slice, the constraints are the linearization of the Einstein equations $G_{0 \mu} = 8 \pi G T_{0 \mu}$ which take the explicit form
\begin{align}\label{Constraints}
\partial_i \partial_j h_{ij} - \partial_i \partial_i h_{jj} &= \frac{\kappa}{2} T_{00} \\
\partial_j \dot{h}_{ij} - \partial_i \dot{h}_{jj} + \partial_i \partial_j h_{0j} - \partial_j \partial_j h_{0i} &= \frac{\kappa}{2} T_{0i},\nonumber
\end{align}
where $i,j = 1,\ldots, D-1$ are spatial indices.
These are the linearized Hamiltonian and diffeomorphism constraints, respectively, and we want to determine whether they can be solved so that the field outside $U$ only depends on the Poincar\'e charges of the matter distribution within $U$.

To see the existence of such solutions, we will show as an example that there are solutions where gravitational lines run from the particles of the distribution to a common origin within $U$, and that an appropriate field extending from this origin then yields a gauge-invariant construction with field exterior to $U$ only depending on the Poincar\'e charges.

We begin with a distribution corresponding to a pointlike particle of energy $p^0$ at rest at a position $\vec r = r \hat r$:
\begin{equation}
T_{00}(\vx) = p^0 \delta^{D-1}(\vx - \vec{r}).
\end{equation}
We can associate to this matter source a gravitational  line solution which runs to the origin at $x=0$, whose nonzero components are given by
\begin{equation}\label{linetoO}
h_{ij}(\vx) = \frac{\kappa}{2} p^0 \delta^\perp(\vx) (r - x) \theta(r - x) \left( \hat r_i \hat r_j - \frac{\delta_{ij}}{D-2}  \right),
\end{equation}
where here $\delta^\perp$ denotes the delta function in the $(D-2)$ directions perpendicular to $\hat r$.
One way to derive this metric is by considering the commutator of the linearized metric operator with $p^0 V_0$, where $V_\mu$ is the gravitational Wilson line dressing of Ref.~\cite{DoGi1}. 
The fact that $h_{ij}$ solves the constraints away from the origin then follows from the fact that the constraints generate diffeomorphisms, and that the dressed operator is diffeomorphism-invariant.
We can also check the constraints \eqref{Constraints} explicitly,
\begin{equation} \label{Wilsonline0}
\partial_i \partial_j h_{ij} - \partial_i \partial_i h_{jj} = \frac{\kappa}{2}  \left[  p^0 \delta^{D-1} (\vx - \vec r) - p^0 \delta^{D-1}(\vx) + p^0 r^i \partial_i \delta^{D-1}(\vx) \right].
\end{equation}
and see that they are satisfied at $\vx = \vec r$ and everywhere else except for the origin.

At the origin, the violation of the Hamiltonian constraint has two terms: one proportional to the energy $p^0$, and the other proportional to the boost charge $p^0 r^i$.
With any collection of charges, we can simply take a linear combination of fields of the form \eqref{Wilsonline0}, and the violation will be proportional to the total energy and total boost charge:
\begin{equation}
P^0 = \sum_A p^0_A, \qquad K^i=\sum_A p^0_A r^i_A\
\end{equation}
where the sum is over particles. 
This violation may be remedied by providing another gravitational field with source at $\vx=0$.  
A simple example is to run a gravitational line, or lines, from $\vx=0$ to infinity.

To construct such a solution, we can again make use of the dressings constructed in Ref.~\cite{DoGi1}.
To cancel the effect of a nonzero total energy, we can choose to concentrate the gravitational field to a line pointing along the positive $z$-axis:
\begin{equation} \label{Hdressing}
h_{ij} = \frac{\kappa}{2} P^0 z\theta(z) \delta^\perp(\vec x) \left( \hat z_i \hat z_j - \frac{\delta_{ij}}{D-2}  \right)
\end{equation}
where now $\perp$ refers to the directions orthogonal to $\hat z$.
This solution violates the constraint at the origin so that it  precisely cancels out the component proportional to $P^0$ in \eqref{Wilsonline0}:
\begin{equation}
\partial_i \partial_j h_{ij} - \partial_i \partial_i h_{jj} = \frac{\kappa}{2} P^0 \delta^{D-1}( \vec x)\ .
\end{equation}
Similarly to cancel out a net boost charge $K^i$, we can add a solution
\begin{equation} \label{Kdressing}
h_{ij} = -\frac{\kappa}{2} \theta(z)K^k\partial_k\left[ z \delta^\perp(\vec x)  \right]\left( \hat z_i \hat z_j - \frac{\delta_{ij}}{D-2}  \right)\ .
\end{equation}

If the source has nonzero spatial momentum, we can solve the linearized diffeomorphism constraint by a similar method.
Consider a distribution with momentum $p^k$ concentrated at the point $\vec r$:
\begin{equation}
T_{0k} = p_k \delta^{D-1} (\vx - \vec r).
\end{equation}
The corresponding linearized solution to \eqref{Constraints} is slightly more complicated, and is given by
\begin{align} \label{Wilsonlinek}
\dot h_{ij} = -\frac{\kappa}{2} \Biggl\{& \left[ p_k \hat r^k \delta^\perp(\vx) +(r-x) p^k \partial_k \delta^\perp(\vx) \right] \theta(r - x)\left( \hat r_i \hat r_j - \frac{\delta_{ij}}{D-2} \right)\nonumber\\ &+  (p^\perp_i \hat r_j + p^\perp_j \hat r_i)\left[\theta(r - x)\delta^\perp(\vx) -\frac{r}{2} \delta^{D-1}(\vx)\right]\Biggr\},
\end{align}
where $p^\perp$ denotes the projection of the momentum perpendicular to $\hat r$.
Once again this solves the constraints away from the origin:
\begin{equation}
\partial_j \dot{h}_{ij} - \partial_i \dot{h}_{jj} 
= \frac{\kappa}{2} \left[ p_i \delta^{D-1} (\vec x - \vec r)  - p_i \delta^{D-1} (\vec x) - \frac{1}{2} (r_i p_j - p_i r_j) \partial_j \delta^{D-1}(\vec x) \right] \ .
\end{equation}
The violation of the constraint equation now has two terms supported at the origin: one proportional to the momentum, and the other proportional to the angular momentum.  

Again, the constraints may be restored by adding a gravitational line from $\vx=0$ to infinity.  
Suppose the configuration has net momentum and angular momentum given by
\begin{equation}
P^i = \sum_A p^i_A, \qquad L^{ij} = \sum_A (r^i p^j - p^i r^j)_A.
\end{equation}
To cancel the net momentum $P^i$, the line may be taken along the positive $z$ axis with the form
\begin{equation} \label{Pdressing}
\dot h_{ij}=\frac{\kappa}{2}\theta(z)\left[ \left( \hat z_i \hat z_j - \frac{ \delta_{ij}}{D-2} \right)\left(P_z\delta^\perp(\vec x) -z P^k\partial_k\delta^\perp(\vec x)\right) +     (P^\perp_i \hat z_j + \hat z_i P^\perp_j) \delta^\perp(\vec x) \right]\ .
\end{equation}
To cancel the net angular momentum $L^{ij}$, we can add
\begin{equation} \label{Ldressing}
\dot h_{ij} = \frac{\kappa}{2} \left[\hat z_{(i} L_{j)}{}^k-\frac{\delta_{ij}}{D-2}L_z{}^k\right]\partial_k\left[\theta(z) \delta^\perp(\vec x)\right]\ .
\end{equation}
The resulting configuration of the gravitational field, consisting of the sum of \eqref{linetoO} and \eqref{Wilsonlinek} for each particle, plus the sum of \eqref{Hdressing}, \eqref{Kdressing}, \eqref{Pdressing}, \eqref{Ldressing} solves the linearized constraints everywhere.

The construction we have just given provides an example of the claimed classical field configuration: for any distribution of matter within $U$, we may, by superposing the  solutions we have just described, find initial data for the metric such that the metric outside $U$ depends just on the Poincar\'e charges of the distribution.  
Of course, there will be different such solutions; for example, we expect that the asymptotic line dressings we have constructed can be replaced by a linearized version of the boosted Kerr solution, which carries all the relevant Poincar\'e charges.\footnote{We can likewise think of this result as a generalization of the Corvino-Schoen gluing theorem\cite{CoSc,ChDe} to the case of initial data with sources, within the linearized theory.  This theorem states that given vacuum initial data, new  initial data may be found that agrees with the original initial data in a compact region, but matches  a standard solution -- taken in \cite{CoSc,ChDe} to be the boosted Kerr solution -- outside a large enough radius.}

\subsection{Implications for classical hair, and the question of quantum split structure}

Since for any matter distribution in $U$, a standard dressing may be chosen that depends just on the Poincar\'e charges, we have a classical version of the split structure described in \eqref{grsplit}, with sectors labeled by those charges.  This also allows an analysis of the role of the BMS charges that is similar to that in QED.  First, all matter configurations in $U$ with the same Poincar\'e charges will have indistinguishable standard fields outside $U$, and so in particular cannot be distinguished by their BMS charges.  However, one may choose different standard fields, by, {\it e.g.}, switching between the line dressings and the boosted Kerr solution.  These {\it will} have different BMS charges.  
Since these fields have the same sources, their difference is a pure radiation field.  Thus, the data in the BMS charges corresponds to the choice of boundary conditions on the radiative part of the field, but not on the matter configuration, and in particular there is a large amount of information in the latter that is asymptotically undetectable in the classical theory.

If this classical statement provides accurate guidance to the quantum theory, that would indicate that information in a given region, aside from the Poincar\'e charges, cannot be measured asymptotically, and in particular that BMS charges don't constrain such information.  But, there are subtleties in the quantum case, and an important question is to what extent the classical construction generalizes, so that  there are charged sectors in the full (for now, linearized) quantum theory that satisfy the condition for split structures, \eqref{grsplit}.
From the quantum generalization of the preceding analysis, we naively expect that the labels in  \eqref{grsplit} correspond to Poincar\'e representations and states within them.  
Analysis of this question is in progress.  
A specific example of quantum states are coherent states created using dressed operators \eqref{phifar} or analogs with different dressings.  Interestingly, as we plan to describe in upcoming work, further information {\it can} be determined about such states by measurements -- external to the region $U$ where the states are supported -- of correlation functions of {\it products} of the Poincar\'e charges.  The situation for more general states is under investigation.  If there are such split structures in the quantum theory, that would appear to tell us that, at least in the perturbative theory, information may be contained within a region without being visible or accessible from outside --  in particular through measurement of soft charges.

\section{Global symmetries}

Regardless of how the preceding question of the existence of split structures in  gravity is answered in the full quantum theory, there is a clear example of the existence of split structures in the perturbative theory, in the presence of matter with global symmetries.  Two important caveats which we will discuss below are that there are questions about the existence of global symmetries in the full nonperturbative theory of gravity, and that if one considers the full nonperturbative theory, global symmetries don't necessarily lead to the type of split structure described above.  Before addressing these, we describe how global symmetries perturbatively lead to split structures.

For a simple example, consider gravity coupled to equal-mass scalar fields $\phi_1$, $\phi_2$, related by an unbroken $U(1)$ symmetry.  Given an open region $U$, let $|\widehat \ue\rangle$ be a split vacuum, and consider dressed operators $A_1$ and $A_2$ built solely from the fields $\phi_1$ or $\phi_2$, respectively.  If these operators are identically constructed and dressed, they will be related by the $U(1)$ symmetry.  We can then for example consider states $|\psi,U_\epsilon\rangle$ and $|\tilde\psi,U_\epsilon\rangle$ that are general linear combinations of $A_1|\widehat \ue\rangle$ and $A_2|\widehat \ue\rangle$.  Given any operator $A'$ outside $\ue$, these will obey a relation \eqref{grsplit} which says that $A'$ cannot distinguish the linear combinations in question.  Put differently, we can think of the states $A_1|\widehat \ue\rangle$ and $A_2|\widehat \ue\rangle$ as realizing a ``qubit" whose state cannot be measured by operators outside $\ue$.  For even more concreteness, $A_1$ and $A_2$ could be taken to be dressed fields $\Phi_1$, $\Phi_2$ of the form \eqref{phifar}, with either of the field operators 
$\phi_1$ or $\phi_2$ dressed by an identical Faraday line, or by another dressing such as the Coulomb dressing of \cite{DoGi1}.

While this example appears illustrative, the status of global symmetries and their charges in gravity is uncertain.\footnote{For one further discussion, see \cite{KLLS}.}  A first reason for this is black hole evaporation, in which the Hawking radiation has no apparent reason to carry out any global charge of the initial black hole.  If this extends to virtual effects, for example virtual black holes, such effects would generically violate any global symmetry.  Alternately, it has been argued that a quantum treatment of gravity may include spacetime wormholes\cite{GiSt}, and if so, global charge can be carried away, and thus violated for our spacetime, by those wormholes.   Finally, if string theory is nature's theory of quantum gravity, it has been argued in \cite{BaDi} using world-sheet methods that there are no continuous global symmetries.  Nevertheless, we regard the question of global symmetries as open, pending better understanding of the nonperturbative structure of gravity, and in particular the explanation of how the theory can be unitary.  It is interesting that global symmetries provide a clear perturbative example of split structure, though the situation potentially changes in the nonperturbative theory, as we discuss next.

\section{Measurement via non-perturbative gravitational translation?}

The preceding section has argued that global symmetries -- should they exist in the full non-perturbative quantum theory -- lead to examples of split structures in the perturbative theory; our earlier discussion raised the more general question of existence of split structures in the perturbative theory.  The present section will explore an important question regarding whether such structures persist in the full nonperturbative theory.

For a concrete example, consider dressed fields $\Phi_1(x)$, $\Phi_2(x)$ like we have just described; if the $U(1)$ symmetry persists nonperturbatively, these lead to states perturbatively indistinguishable outside a neighborhood $U$ containing $x$.   However,  this is not so {\it nonperturbatively}.  This arises because in gravity, translation operators can be formed from the fields {\it outside} $U$.  Specifically, spatial translations are generated by the ADM momentum
\beq\label{admmom}
P_i = -\frac{2}{\kappa} \oint r^2 d^2\Omega {\hat r}^j\left(\dot h_{ij}-\dot h_{kk} \delta_{ij}+\partial_i h_{0j}-\partial_j h_{0i}\right)
\eeq
which is supported at infinity and translates fields through its action on the dressing\cite{DoGi1}.  So, for example, if $U$ is centered on $x=0$, the operator
\beq
e^{-i a^i P_i} \Phi_1(0) e^{i a^i P_i} = \Phi_1(a) 
\eeq
can be translated to an $a$ arbitrarily far outside $U$.  This in principle allows us to distinguish the states $\Phi_1(0) |\widehat \ue\rangle$ and $\Phi_2(0) |\widehat \ue\rangle$ via observables ``outside $U$:"  for $x'$ outside $\ue$, there are clearly operators $A(x')$ such that a combination of operators completely localized outside $\ue$ distinguishes these states:
\beq\label{opshift}
\langle \widehat \ue |\Phi_1(0)e^{i a^i P_i}A(x')e^{-i a^i P_i}\Phi_1(0)|\widehat \ue\rangle \neq \langle \widehat \ue |\Phi_2(0)e^{i a^i P_i}A(x')e^{-i a^i P_i}\Phi_2(0)|\widehat \ue\rangle\ .
\eeq
Examples for $A(x')$ are the individual currents $\phi_\alpha\partial \phi_\alpha$, with $\alpha=1$ or 2, or their dressed versions.  One might even anticipate that this kind of construction could distinguish states inside a black hole -- effectively the translation operator $P_i$ can be viewed as translating the observable into the black hole, and then back out, or translating the black hole interior to the observation point $x'$, then back.

However, this argument that one can observe information in a region from outside that region relies on the  non-perturbative structure of gravity.  Indeed, one first indicator of this is the $1/\kappa$ in \eqref{admmom}, showing that the translation operator is not perturbatively defined.  Because of this, $P_i$ acting on a dressing $V^\mu_{(1)}$ that is first order in $\kappa$, like those given in \cite{DoGi1}, gives a {\it zeroth} order shift:
\beq
e^{-i a^i P_i} V^j_{(1)}(x) e^{i a^i P_i} = V^j_{(1)}(x) + a^i\ ,
\eeq
so
\beq
e^{-i a^i P_i} \phi(x+V_{(1)}(x)) e^{i a^i P_i} = \phi(x+a+V_{(1)}(x)) \neq \phi(x+a+V_{(1)}(x+a))\ .
\eeq
To shift the operators $\Phi(x)$ a finite distance, as needed to move them outside the original region $U$, as in \eqref{opshift}, one needs the full non-perturbative (all-orders) dressing $V^\mu(x)$.

The all-order quantum operator $\Phi(x)=\phi(x+V(x))$  is found by solving the nonperturbative version of the constraint equations $G_{0\mu}=8\pi G T_{0\mu}$ to all orders in the quantum theory.  This is equivalent to having the full non-perturbative evolution of the system, since the $\mu=0$ constraint contains time evolution.  Thus, while it appears that formally we have the ability to determine the information content of a region from outside that region in gravity, that rests on the assumption that we have already found the full non-perturbative unitary evolution.  Of course, the latter problem is particularly difficult, given the challenges to unitarity inherent in non-perturbative gravitational phenomena, such as black hole formation.  But, once we have solved the full problem of non-unitary quantum evolution, it appears that we could have such ``non-local" access to information, and in this sense non-perturbative departures from split structure.  However, given this non-triviality it is not clear how such statements of a possible origin of ``gravitational holography" \cite{Maroholo1,Maroholo2} can lead to an explanation for the mechanism behind such unitary evolution.  Put differently, while there appears to be a link between nonlocality and unitary evolution in gravity, it is not yet clear that we have a more basic explanation of the reasons for either of them, that may be used to deduce the other.
\vskip.1in
As this paper was being finalized, a related paper \cite{Bousso:2017dny} appeared, making similar points about soft hair not constraining information content of a region.

\section{Acknowledgements}

We wish to thank 
C. Fewster, J. Hartle, S. Hollands, D. Marolf, A. Strominger, and J. Tener for helpful discussions.
This work was supported in part by the U.S. DOE under Contract No. 
DE-SC0011702,  Foundational Questions Institute (fqxi.org) 
Grant No. FQXi-RFP-1507, the University of California, and by National Science Foundation Grant No. NSF PHY11-25915 to the Kavli Institute of Theoretical Physics, whose hospitality is gratefully acknowledged.

\bibliographystyle{utphys}
\bibliography{QG-sub}

\end{document}